\documentclass[aps,pra,twocolumn,amsmath,amssymb,superscriptaddress,longbibliography]{revtex4-2}
\usepackage{amssymb}

\usepackage{url} 
\usepackage{amsmath}
\usepackage{amssymb}
\usepackage{amsthm}
\usepackage{fontenc}
\usepackage{graphicx}
\usepackage{xcolor}
\usepackage{textcomp}
\usepackage{epstopdf}
\usepackage{braket}
\usepackage{mathtools}
\usepackage{amsmath}
\usepackage{dcolumn}
\usepackage{multirow}
\usepackage{gensymb}
\usepackage{bbm}
\usepackage{bm}

\begin{document}

\title{Universality of moiré physics in collapsed chiral carbon nanotubes}
\author{Olga Arroyo-Gasc\'on}
\email{o.arroyo.gascon@csic.es}
\affiliation{Instituto de Ciencia de Materiales de Madrid, Consejo Superior de Investigaciones Cient\'{\i}ficas, Sor Juana In\'es de la Cruz 3, 28049 Madrid, Spain}
\affiliation{GISC, Departamento de F\'{\i}sica de Materiales, Universidad Complutense, E-28040 Madrid, Spain}
\author{Ricardo Fern\'andez-Perea}
\affiliation{Instituto de Estructura de la Materia (IEM), CSIC, Serrano 123, E-28006 Madrid, Spain}
\author{Eric Su\'arez Morell}
\affiliation{Departamento de F\'isica, Universidad T\'ecnica Federico Santa Mar\'ia, Casilla 110-V, Valpara\'iso, Chile}
\author{Carlos Cabrillo}
\affiliation{Instituto de Estructura de la Materia (IEM), CSIC, Serrano 123, E-28006 Madrid, Spain}
\author{Leonor Chico}
\affiliation{GISC, Departamento de F\'{\i}sica de Materiales, Universidad Complutense, E-28040 Madrid, Spain}

\date{\today}

\begin{abstract}
We report the existence of moiré patterns and magic angle physics in all families of chiral collapsed carbon nanotubes. 
A detailed study of the electronic structure of all types of chiral nanotubes, previously collapsed via molecular dynamics,
has been performed. 
We find that each family possesses a unique geometry and moiré disposition, as well as a characteristic number of flat bands. 
Remarkably, 
 all kinds of nanotubes behave the same with respect to magic angle tuning, showing a monotonic behavior that gives rise to magic angles  
in full agreement with those of twisted bilayer graphene. 
 Therefore, magic angle behavior is universally found in chiral collapsed nanotubes with a small chiral angle, giving rise to moiré patterns. 
 Our approach comprises first-principles and semi-empirical calculations of the band structure, density of states and spatial distribution of the localized states signaled by flat bands.
\end{abstract}

\maketitle

\section{Introduction}

In recent years, layered materials showing moiré patterns have stood as a platform for
the exploration of new physics \cite{andrei_marvels_2021,devakul_magic_2021,naik_ultraflatbands_2018,cao_correlated_2018-1,cao_unconventional_2018-2,nayak_probing_2017,tran_evidence_2019,wu_hubbard_2018,leon_tuning_2022}. Graphene-based materials, such as twisted bilayer graphene (TBG) and transition metal dichalcogenides, constitute two families of great interest.
In particular, layered graphene structures have been shown to host electronic correlations, superconductivity and nontrivial topological phases when arranged at a small rotation angle, the so-called magic angle, which is close to 
$1\degree$ for TBG \cite{cao_correlated_2018-1,cao_unconventional_2018-2,suarez_morell_flat_2010-1,song_all_2019,serlin_intrinsic_2020,tarnopolsky_origin_2019,lisi_observation_2021}.
 
The quest for novel materials which could enlighten these new phenomena has even extended to one-dimensional systems, where many-body effects have been widely described. Flat bands can be found in some one-dimensional materials
 \cite{huda_designer_2020,kennes_one-dimensional_2020,arroyo-gascon_one-dimensional_2020,zhao_interlayer_2022}; for instance, carbon nanotubes (CNT) can show moiré patterns in their double-walled and multi-walled form \cite{bonnet_charge_2016-1,zhao_observation_2020-1,koshino_incommensurate_2015-1}, and single-walled tubes also display moirés when collapsed \cite{arroyo-gascon_one-dimensional_2020}. For the latter, flat bands with an even smaller energy span than TBG and sharp densities of states ensue along with these one-dimensional patterns. As in TBG, these features depend on 
 the rotational angle, which is related to a corresponding chiral angle in collapsed CNTs.

The electronic structure of carbon nanotubes is varied, 
comprising 
metallic and semiconducting tubes.
In metallic 
CNTs, which host Dirac fermions as in graphene, the Dirac point can be located at the center of the Brillouin zone  (the $\Gamma$ point) or at 2/3 of the $\Gamma$-X line \cite{dresselhaus_group_2008-1}. 
 We denote them as $\Gamma$-metals and 2/3-metals, respectively \cite{arroyo-gascon_one-dimensional_2020}. Chiral metallic CNTs can belong to these two 
groups, although a small gap develops due to curvature effects. 
In a previous article, we found flat bands and highly localized states 
in the AA regions (zones of AA stacking) of the moiré patterns formed in collapsed 2/3-metal nanotubes with small chiral angles \cite{arroyo-gascon_one-dimensional_2020}. In that work we chose to explore 2/3-metallic CNTs because of their similarity to graphene: they present two bands with linear dispersion crossing at 2/3 of the positive part of their Brillouin zone, being the one-dimensional (1D) analogues of the Dirac cones in graphene. 
However,
the behavior of collapsed chiral $\Gamma$-metals and semiconducting tubes is yet unknown;
further analysis is needed to 
elucidate whether  
 magic angle 
physics also pertains 
 to the rest of families of chiral tubes. 

In order to obtain CNTs that are stable upon collapse 
with sizable moirés at small chiral angles, 
tubes with diameters above 40 \AA, 
often involving a high number of atoms per unit cell, are needed \cite{chopra_fully_1995,benedict_microscopic_1998,tang_collapse_2005,gao_energetics_1998,liu_molecular_2004,elliott_collapse_2004,zhang_closed-edged_2012,he_precise_2014,impellizzeri_pc_2019,he_review_2019}. 
Note that for free-standing collapsed nanotubes the threshold is higher, 51 \AA\ \cite{he_precise_2014,impellizzeri_pc_2019}, although deposition on a substrate can lower this limit substantially \cite{he_review_2019}. 
The search for $\Gamma$-metal and semiconducting tubes which fulfill these conditions leads to nanotubes showing several  
AA regions per unit cell, which is intrinsically related to the symmetry operations and the number of localized states in the structure.

In this work, we show that moiré physics occurs for all collapsed chiral carbon nanotubes close to the magic angle, regardless of their metallic or semiconducting behavior. 
We assess several criteria usually 
employed 
to describe magic angle physics: the appearance of flat bands and the reduction of the Fermi velocity, sharp peaks in the density of states, and real-space electronic localization via local density of states \cite{arroyo-gascon_one-dimensional_2020,cao_correlated_2018-1,andrei_marvels_2021,trambly_de_laissardiere_localization_2010}. We study the interplay of these criteria and discuss the most suitable one for our system. Analyzing these benchmarks, we perform an exhaustive description of the electronic structure of a range of collapsed chiral nanotubes belonging to each family: semiconducting, 2/3-metals and $\Gamma$-metals.

We find a magic angle very close to that of TBG for the three families of CNTs, namely, $1.12\degree$ for 2/3-metals, and $1.11\degree$ for semiconducting and $\Gamma$-metals. 
Moreover, a homogeneous behavior extends to all tubes when their 
moir\' e angle is small enough, regardless of their specific family or symmetries. Our findings imply that the experimental observation of one-dimensional moiré physics in CNTs might be easier than previously expected.

\section{Geometry and symmetry} \label{sec:geometry}

Let us briefly recall the standard notation in nanotube physics. CNTs are identified by its circumference vector $\bm{C_h}$ on an unrolled graphene sheet. 
The coordinates of $\bm{C_h}$ in the graphene basis, $(n,m)$,  are customarily used to label CNTs. Considering a one-orbital model and leaving out curvature effects, if $n-m$ is a multiple of 3, the tube is metallic \cite{saito_electronic_1992-2,hamada_new_1992-1,saito_electronic_1992-3}.
Another important magnitude is the chiral angle of the CNT, $\theta_{NT}$, which is spanned between the $(n,0)$ direction and $\bm{C_h}$. 
Upon collapsing a chiral carbon nanotube, a moiré pattern can emerge, which is directly related to the CNT chiral angle $\theta_{NT}$. 
Thus, a moiré angle $\theta_M$ can be defined as the relative rotation angle between the two flattened parts of the CNT, just as in TBG. An analysis of the geometry of the unrolled unit cell yields $\theta_M=2\theta_{NT}$ \cite{arroyo-gascon_one-dimensional_2020}, so that in the following tubes will be labeled by the angle $\theta_M$.

\begin{figure}[h!]
 \begin{center}
\includegraphics[trim={1.5cm 0 0 0},clip,width=1.05\columnwidth]{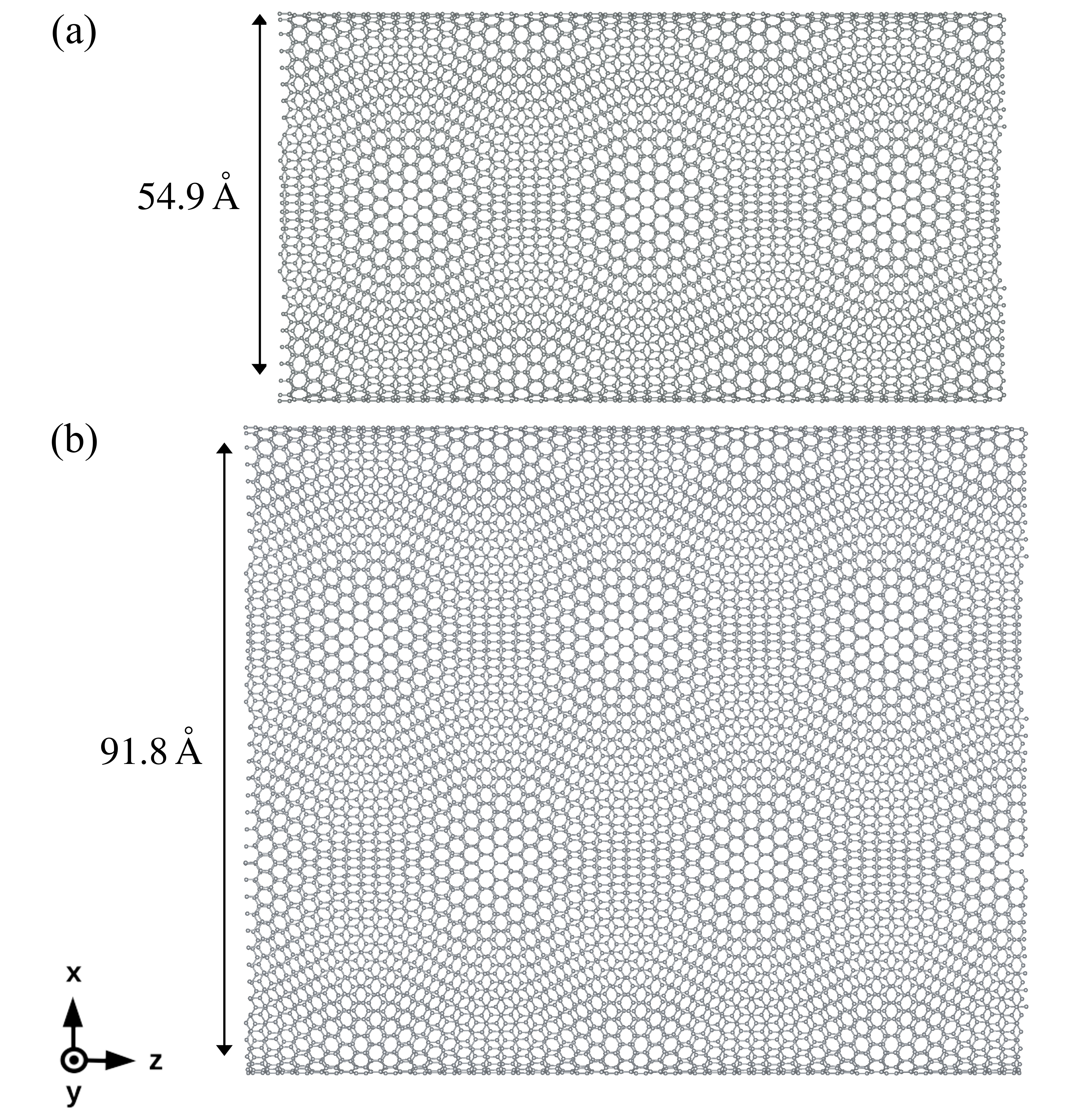} 
\caption{Top view of the unit cells of the (a) collapsed semiconducting (48,2) and (b) $\Gamma$-metal (78,3) nanotubes. The number and placement of the AA regions differ and depend on the symmetry operations of the original cylindrical tube.}
\label{fig1}
\end{center}
\end{figure}

Addressing a wider range of nanotube types implies that new tube symmetries may be involved. Since the geometry of each tube is related to its symmetry operations, the existence and arrangement of moiré patterns once the structure is collapsed can be predicted. Line groups describe the symmetry operations of one-dimensional systems, such as CNTs. Specifically, chiral cylindrical nanotubes belong to the fifth line group
and comprise screw-axis and isogonal point group $D_p$ symmetries, where $p$ is the greatest common divisor of the nanotube indices $n$ and $m$ \cite{damnjanovic_line_2010}. Dihedral groups include rotations and often reflections; 
 in particular, a rotational $C_{2 q}$ 
 symmetry operation, where $q$ is an integer, is convenient, so  
 to obtain two equivalent graphene regions on the unrolled unit cell, that will 
 constitute the two coupled ``layers" of the collapsed nanotube 
 yielding a full AA moir\'e 
 area. In fact, this is the simplest symmetry that yields full 
 AA moiré patterns once the tube is collapsed. This rotational symmetry can be achieved by choosing nanotubes with indices $(2m,2)$. We have followed this method for 2/3-metallic and semiconducting nanotubes; despite having the same greatest common divisor $p$, 2/3-metals and semiconducting tubes might not have the same symmetry operations and exhibit different moiré patterns. In fact, as it is shown in Figure~\ref{fig1}(a), a $C_2$ symmetry not only can render one centered AA moiré region per unit cell as in 2/3-metals, but three.
  
As for $\Gamma$-metallic CNTs, they usually have more atoms per unit cell than 
semiconducting chiral CNTs with 
  similar chiral angles. Suitable nanotubes should be at least 
  40 \AA-wide so that they are stable once collapsed. The smallest tubes that fulfill the aforementioned diameter condition and have a moiré angle close to the magic angle in TBG have indices $(3m,3)$, thus showing $C_3$ symmetry. This results on 
 a zigzag arrangement of the AA moiré regions with respect to the tube axis 
(see Figure \ref{fig1}(b)), 
in contrast 
to the linear disposition achieved for 2/3-metallic and semiconducting tubes. The minimum number of 
AA regions per unit cell for these tubes is now six (taking into account 
half AA 
moiré spots). Therefore, we show that not only $C_{2q}$-symmetric tubes are suitable for moiré engineering, and that 
sizable AA regions can appear in all families of CNTs. 

The appearance of AA-stacked moiré patterns in the center of collapsed chiral tubes, such as in TBG, motivates the search for fragile topological phases in CNTs, that have been predicted and found in TBG \cite{song_all_2019,serlin_intrinsic_2020} and other systems displaying flat bands \cite{skurativska_flat_2021}. For instance, edge states in cylindrical CNTs can be classified attending to a topological invariant \cite{okuyama_topological_2019}. However, a group theory analysis of the band representations following \cite{bradlyn_topological_2017} does not render fragile topology \cite{milosevic_elementary_2020}. Regarding collapsed CNTs, the molecular dynamics process can break the rotational symmetries of the tubes, so that symmetry-protected fragile topological states, such as $C_2T$ in TBG \cite{song_all_2019}, are not feasible in our case. Nevertheless, model Hamiltonians used to describe nontrivial topology in TBG could be applied to the central part of our tubes, where the AA regions appear, since $C_2$ symmetry is locally present there.


\section{Methods}

Collapsed nanotubes are simulated by means of molecular dynamics calculations, resorting to the Large-scale Atomic/Molecular Massively Parallel Simulation (LAMMPS) package \cite{thompson_lammps_2022}. This allows us to have a reliable description of the structure. In fact, at low angles we reproduce the corrugations that are known to appear in TBG \cite{wijk_relaxation_2015-3} in the flattened, bilayer-like portion of the nanotubes. An Adaptive Intermolecular Reactive Empirical Bond Order (AIREBO) \cite{stuart_reactive_2000} potential models the interactions between carbon atoms. Periodic conditions are applied with supercells wide enough to avoid interaction between nanotube replicas. A detailed description of the methods used in this work can be found in the Supporting Information of \cite{arroyo-gascon_one-dimensional_2020}. 
Figure \ref{fig1} shows two examples of converged final geometries. 

The band structure of the tubes is then obtained by means of a tight-binding model derived from that presented by Nam and Koshino \cite{nam_lattice_2017}:
\begin{equation}
H= -\sum_{i,j} t({\mathbf R}_i - {\mathbf R}_j)  \ket{{\mathbf R}_i}   \bra{ {\mathbf R}_j } + {\rm H. c.},
\end{equation}
where ${\mathbf R}_i$ is the position of the atom $i$, $ \ket{{\mathbf R}_i}$ is the wavefunction at $i$, and $t({\mathbf R})$ is the hopping between atoms $i$ and $j$:
\begin{equation}
-t({\mathbf R}) = V_{pp\pi} (R)  \left[ 1-\left( \frac{ {\mathbf R} \cdot  \mathbf{e}_y } {R} \right)^2 \right] +  V_{pp\sigma} (R) \left( \frac{ {\mathbf R} \cdot  \mathbf{e}_y } {R} \right)^2
\end{equation}
The explicit expression of the hopping parameters $V_{pp\pi}(R)$ and $V_{pp\sigma} (R)$ is detailed in the Supporting Information of \cite{arroyo-gascon_one-dimensional_2020}.
After the molecular dynamics calculation, the tubes end up with a flattened central region where moirés appear (see Figure~1), as well as two narrow lobular regions at the extremes of the flattened part. Both $V_{pp\pi}(R)$ and $V_{pp\sigma}(R)$ contributions are taken into account for the flat part. Only the $V_{pp\pi}(R)$ part of the Hamiltonian is applied to the regions of the lobes, since we do not need to assess the effect of corrugations
therein.

\begin{figure}[h!]
\begin{center}
\includegraphics[width=1.05\columnwidth]{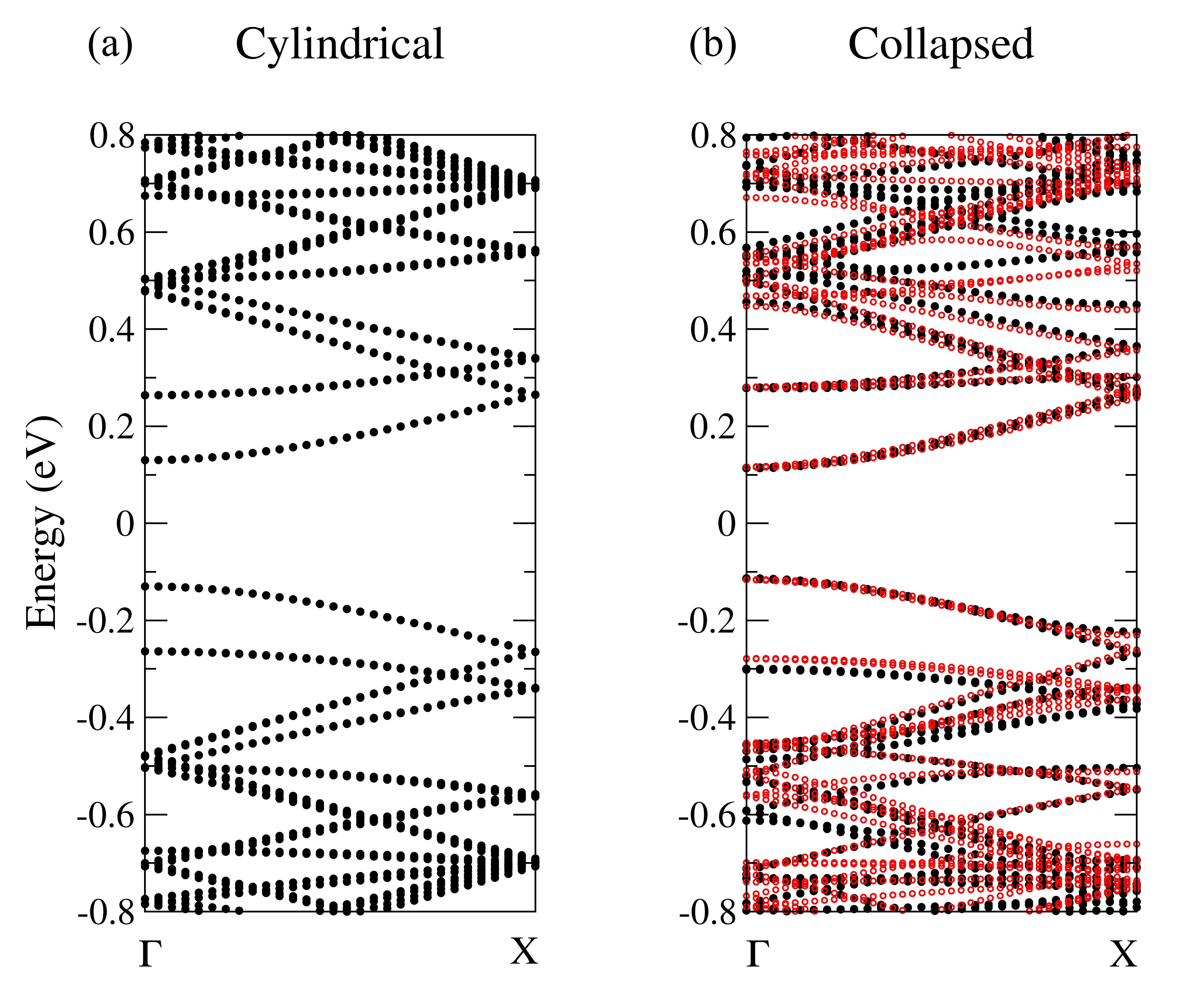}
\caption{Band structures of the (36,2) CNT with $\theta_M=5.36\degree$:  (a) for the cylindrical geometry, obtained with the tight-binding model; (b) for the collapsed structure, obtained with both tight-binding (black) and DFT (red) calculations.}
\label{fig2}
\end{center}
\end{figure}

The tight-binding Hamiltonian allows us to model nanotubes which would be computationally out of reach within a first-principles approach. In addition, it is validated against DFT calculations of the electronic bands of a relatively small semiconducting tube, namely the (36,2), employing the SIESTA code within the generalized gradient approximation (GGA) using the Perdew-Burke-Ernzerhof (PBE) parameterization  \cite{soler_siesta_2002-1,troullier_efficient_1991,perdew_generalized_1996} and a $1\times 1\times 4$ Monkhorst-Pack grid sampling of the Brillouin zone. Even though the GGA-PBE formalism in general underestimates the band gap for semiconductors, a hybrid functional approach is not feasible since we are dealing with thousands of atoms per unit cell. Moreover, we have found that most collapsed CNTs near the magic angle are metallic or have a negligible band gap. We are therefore able to compare the first-principles and tight-binding electronic structures of the collapsed tube, 
where the atomic coordinates are taken from the relaxed nanotube atomic configuration previously obtained by molecular dynamics simulations. Figure \ref{fig2} 
illustrates the agreement between tight-binding (black) and DFT (red) calculations; it is especially important to reflect properly the behavior of the central bands nearest to the Fermi energy, since they will be the most affected upon collapse. Note that the bands of the (36,2) tube, which has 2744 atoms per unit cell, only show a slight flattening since the moiré angle is $\theta_M=5.36\degree$, which has to be compared with the magic angle in TBG, $\sim 1.1\degree$. 

\section{Benchmarks of magic angle physics in semiconducting and $\Gamma$-metal CNTs}
\subsection{Semiconducting nanotubes} \label{sec:semiconducting}

\begin{figure*}[ht!]
\begin{center}
\includegraphics[trim={0cm 0 0cm 0},clip,width=0.8\textwidth]{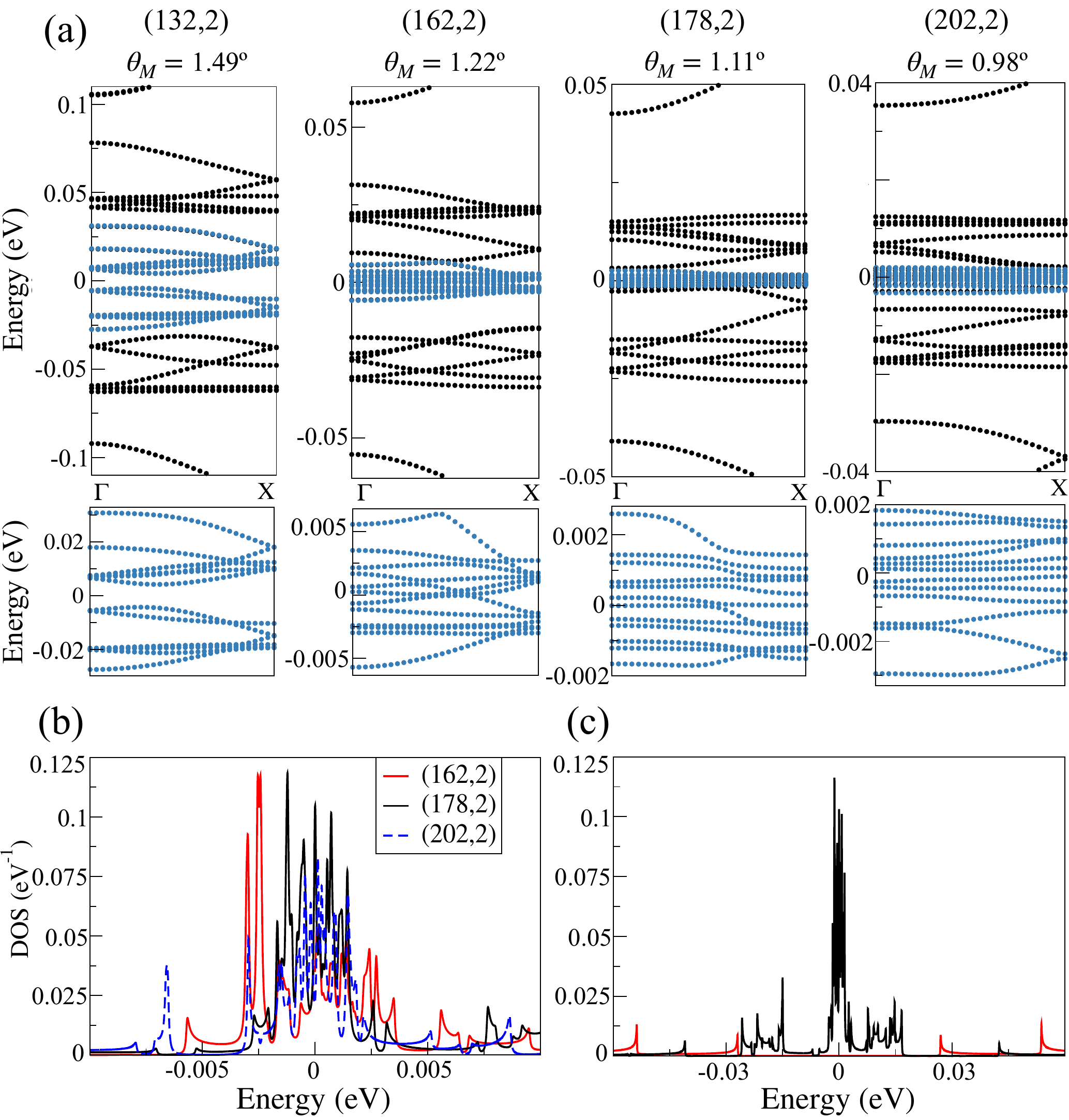}
\caption{(a) Band structures of four semiconducting collapsed tubes; the innermost 12 bands are highlighted in blue and presented in zoom-ins in the bottom panel. Notice the different energy scales. (b) Density of states of the largest three semiconducting tubes in panel (a). (c) Comparison between the collapsed (black) and cylindrical (red) DOS for the (178,2) tube, which displays the maximum peak height among the depicted tubes.}
\label{fig3}
\end{center}
\end{figure*}

In order to extend the flat band picture from 2/3-metal CNTs to all kinds of tubes, the first family we analyze are semiconducting CNTs. As stated in Section \ref{sec:geometry}, we choose these tubes to present a $C_2$ symmetry, which produces three AA regions per unit cell, located 
along the nanotube axis (see Figure \ref{fig1}(a)). Recall that 2/3-metallic CNTs have been shown to host a set of eight flat bands near the neutrality point along with an inner subset of 4 notably flat bands,
and a single AA region per unit cell \cite{arroyo-gascon_one-dimensional_2020}. 
Since the number of AA regions per unit cell has now increased, a higher number of bands are expected to be affected by the collapse. 
Besides, semiconducting tubes fulfilling our requirement also show a higher number of atoms per unit cell 
compared to 2/3-metal CNTs with a similar moir\'e angle.

\begin{figure}[h!]
\begin{center}
\includegraphics[width=0.42\textwidth]{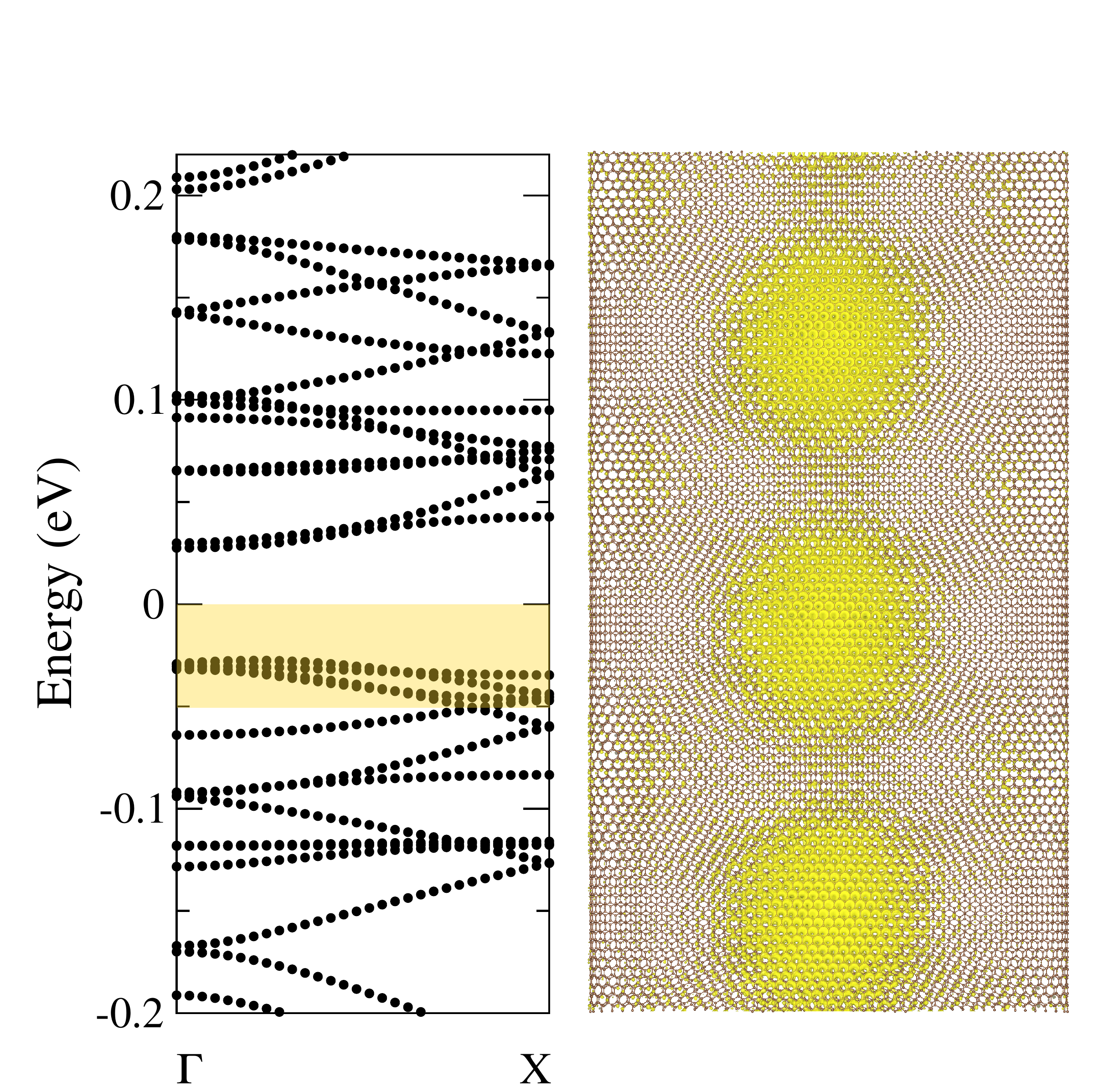}
\caption{Band structure and top view of the (94,2) collapsed semiconducting nanotube. The LDOS of the marked bands is highlighted in yellow.}
\label{fig5}
\end{center}
\end{figure}

Figure \ref{fig3}(a) clearly depicts a progressive flattening and isolation of the central bands of several tubes as the moiré angle decreases, ranging from $1.49\degree$ to $0.98\degree$. This central set now encompasses 24 bands, instead of the 8 bands found in 2/3-metals. 
Remarkably, and as in 2/3-metals, a subset of several extremely flat bands is found in the semiconducting tubes. 
For 2/3-metals, the innermost 4-band set is most flattened; for semiconducting tubes, the number of flat bands in this set increases to 12 and are distinguishable starting from the (162,2) tube in Figure~\ref{fig3}(a). Hence, the number of flattened bands in semiconductors is triple than in 2/3-metals; this is ultimately related to the number of AA regions per unit cell, which is also triple.

Semiconducting tubes can be divided in two groups depending on whether the relation between their indices $n-m=3l \pm 1$, where $l$ is an integer; for instance, the (132,2) and (162,2) tubes belong to the $+1$ group whereas the (172,2) and (202,2) tubes belong to the $-1$ subfamily. They have been shown to display different optical and electronic properties \cite{kataura_optical_1999,chico_curvature-induced_2009,charlier_electronic_2007}, so it is reasonable to ask whether such differences may arise with respect to moir\'e physics. However, observing Fig.~\ref{fig3}(a) we find that the $\pm 1$ classification does not have a significant effect on the flattening of the central bands of the tubes. 

\begin{figure*}[ht!]
\begin{center}
\includegraphics[width=0.8\textwidth]{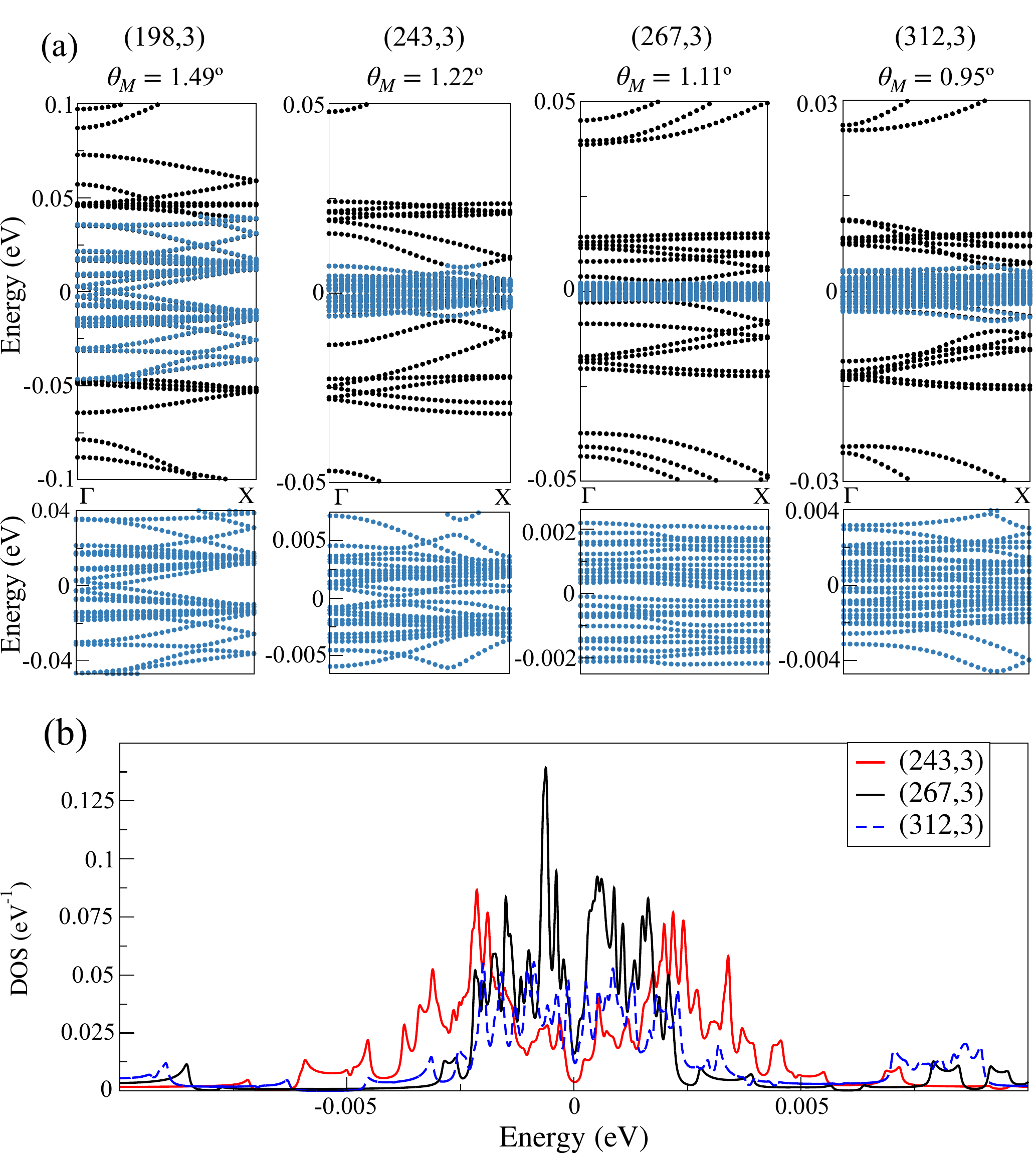}
\caption{(a) Band structures of four collapsed $\Gamma$-metal nanotubes. Mind the different energy scales. (b) DOS of the largest three nanotubes depicted in panel (a).}
\label{fig6}
\end{center}
\end{figure*}

We previously found \cite{arroyo-gascon_one-dimensional_2020} that in 2/3-metals, as in TBG, flat bands 
give rise to 
sharp signals of electronic localization in the density of states (DOS). This is a consequence of the 
proportionality of the DOS with the inverse of the norm of the electronic velocity integrated over the energy surface. In our 1D case, such a relation 
is inversely proportional to the electronic speed, so that the DOS per atom reads,
\begin{equation}
\label{doseq}
\rho(\varepsilon)= \frac{1}{\rho_{A} D_{cnt }}\sum_{n}\frac{1}{\hbar \, |v_{n}(k)|_{k=k_{n}(\varepsilon)}} \,,
\end{equation}
\noindent
where $n$ labels the electronic bands, $v_{n}(k)$ is the corresponding electron velocity, $\rho_{A}$ is the carbon areal density of graphene, and $D_{cnt}$ the diameter of the CNT. Expression (\ref{doseq}), since $\hbar \, |v_{n}(k)|$ is the derivative of the band dispersion relation, neatly illustrates that flattening of the bands gives rise to high narrow peaks in the DOS. 
The presence of higher and narrower peaks is a signature of higher localization, and thus, of a larger probability of strongly correlated electronic behavior. Comparing the DOS peak heights thus serves as a criterion for the potential for strong electronic correlations among diverse nanotubes. 
Additionally, Eq.\ (\ref{doseq}) also implies that for a given resolution the DOS is proportional to the inverse of the electronic velocity averaged over the resolution function. But notice that if, alternatively, we choose as localization criterion the vanishing of electronic velocities in absolute terms, the heights we must compare are those of the DOS multiplied by the corresponding nanotube diameter. All the DOS presented in this work have been computed with a 50 $\mu$eV Lorentzian resolution (energy broadening). 

Figure~\ref{fig3}(b) shows the DOS for the largest semiconducting tubes in Figure~\ref{fig3}, computed from the dispersion relations. 
 Due to the flattening, these tubes become metallic and sharp peaks appear near the neutrality point, whereas a small gap is observed for smaller tubes. As the moiré angle decreases, the peaks pack together around the Fermi level. Moreover, the innermost 12-band set is also separated from the rest of the spectrum. The maximum DOS peak height is reached for the (178,2) tube, with an angle $\theta_M=1.11\degree$. Multiplying by the corresponding diameters does not change the qualitative picture but enhances the preponderance of the (178,2) tube.

The effect of collapse on the DOS is 
illustrated in Figure~\ref{fig3}(c), which spans the 16 innermost bands of the (178,2) tube, both cylindrical and collapsed. The former spaced bands in the cylindrical structure gather around the Fermi level upon collapse and give rise to pronounced peaks that evidence the high localization of these states.  As the LDOS for the smaller (94,2) CNT displayed in Figure~\ref{fig5} confirms, the states are also localized in the AA regions. These three AA-stacked regions show equal enhancement of the DOS, consistently with the triple number of flat bands present in these tubes with respect to those found in 2/3-metals. Notice also that electron-hole symmetry breaks when the tube is flattened due to the interlayer hopping in our model. 

Even though the DOS criterion highlights the (178,2) tube, analyzing the energy span of the tubes depicted in Figure~\ref{fig3} we find that the energy spans for the (178,2) tube are 0.2 meV and 0.52 meV for the lower conduction and highest valence band respectively (conduction band and valence band from now on), larger than those of the (202,2) tube (0.04 meV and 0.15 meV respectively). 
Therefore, for semiconducting nanotubes, the DOS and the band span criteria point to different nanotubes.


\subsection{$\Gamma$-metal nanotubes}

$\Gamma$-metal tubes are similar to 2/3-metals, 
in that they both display a Dirac-like dispersion that is renormalized upon collapse.
 However, the smaller suitable $\Gamma$-metals show six
AA regions per unit cell instead of one (see Figure~\ref{fig1}(b)). Although
the number of 
AA regions per unit cell is six times that of the 2/3-metals, 
that does not necessarily mean sextuple localized states: comparing Figures \ref{fig1}(a) and (b), it follows that the symmetry of $\Gamma$-tubes is different to that of 2/3-metallic and semiconducting tubes
in a way that precludes the tube structure to be approximated 
by consecutive copies of 
single-AA-region cells, as it is the case with semiconducting CNTs.

\begin{figure}[h!]
\begin{center}
\includegraphics[width=1.01\columnwidth]{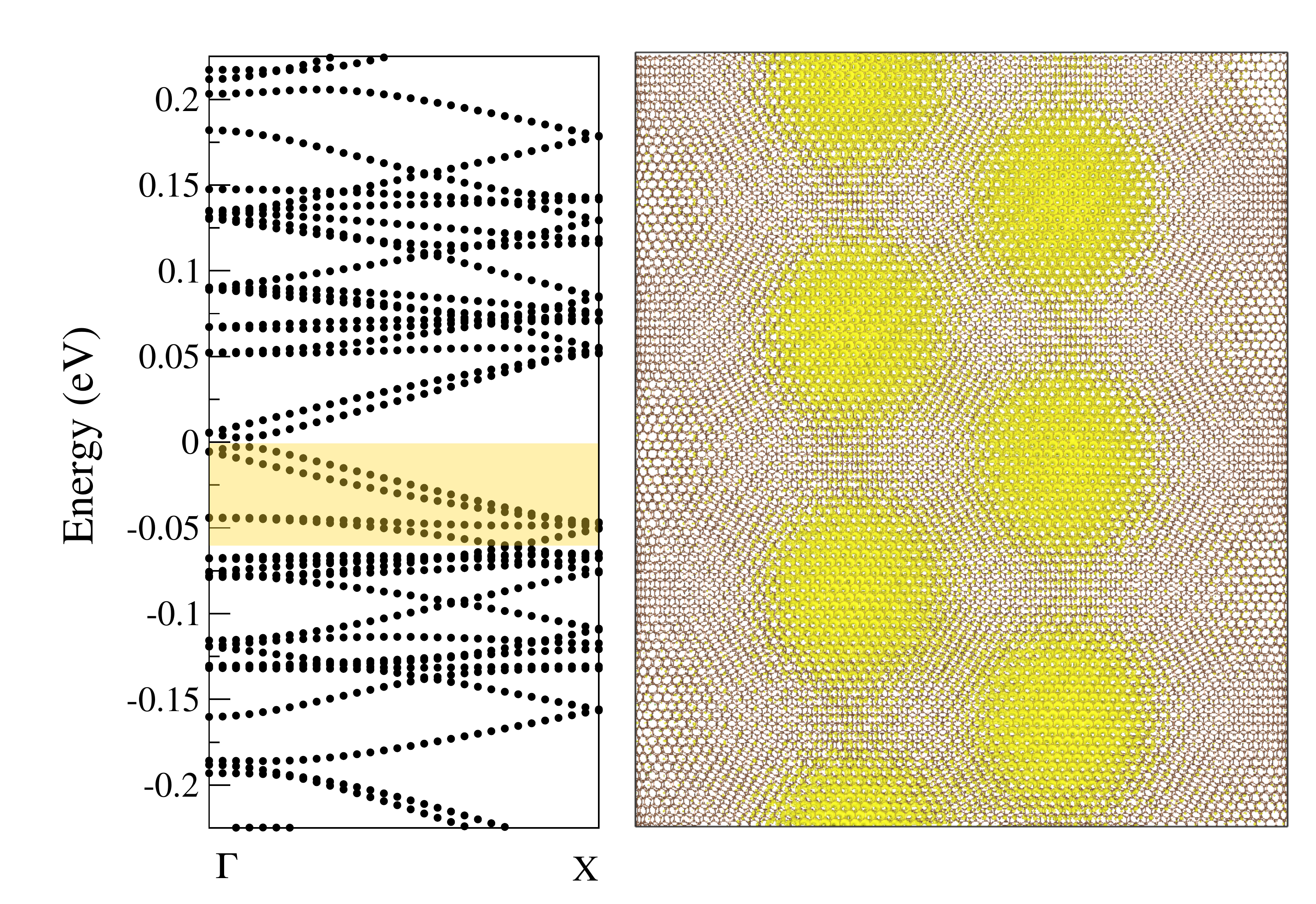}
\caption{Band structure and top view of the (132,3) $\Gamma$-metal nanotube. The LDOS of the marked bands is highlighted in yellow. }
\label{fig7}
\end{center}
\end{figure}

As depicted in Figure~\ref{fig6}(a), the Dirac crossing is slightly displaced away from the $\Gamma$ point for the collapsed tubes. This Figure is equivalent to Figure~\ref{fig3} (a) but for $\Gamma$-metals, with moiré angles between $1.49\degree$ and $0.95\degree$. The bands constituting the Dirac cone are only distinguishable in the leftmost panel. Overall, a set of 36 bands are progressively isolated and flattened, with an inner set of 24 very flat bands. This subset has certainly sextuple bands that the equivalent set in 2/3-metals, in spite of the above-mentioned different symmetry. And indeed, the states are again equally localized at the six AA regions, as Figure~\ref{fig7} illustrates.
Therefore, one can establish for all kinds of nanotubes a direct proportionality between the number of AA regions and the number of flat bands 
around the neutrality point that appear at small chiral angles.

Figure~\ref{fig6}(b) shows the corresponding DOS of the three largest tubes. The most prominent peak appears in the (267,3) case with a moiré angle of $\theta_M=1.11\degree$. The cluster of narrowest peaks show this time a bimodal structure around the Fermi level, i.e., the bands nearest to Fermi are not particularly flat, something that can be gauged also in the band plots (zooms in Figure~\ref{fig6}(a)).

As in the semiconductors, the smallest energy span among $\Gamma$-metals does not correspond to the case with the most prominent DOS peak. In particular, the (312,3) tube (moiré angle $\theta_M=0.95\degree$) has 
0.16 meV and 0.06 meV energy spans for conduction and the valence bands respectively, to be compared with 0.27 meV and 0.16 meV in the case of the (267,3) tube.
 
\subsection{Global discussion and analysis for all CNT types} 
\begin{figure}[h!]
\begin{center}
\includegraphics[width=\columnwidth]{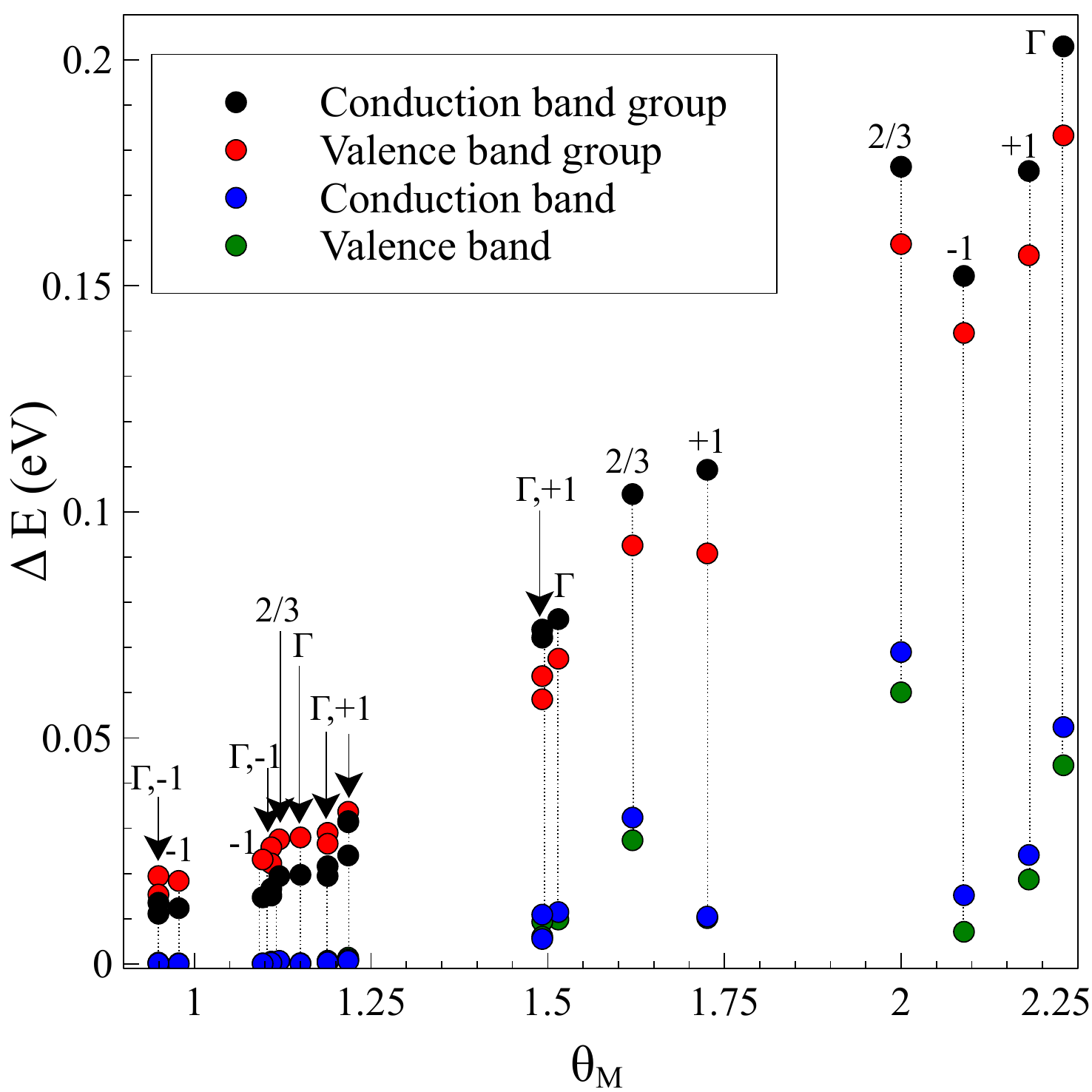}
\caption{Energy spans for the sets of flat bands in all kinds of CNTs. $2/3$ and $\Gamma$-metals are labeled $2/3$ and $\Gamma$, respectively, whereas the two subfamilies of semiconductors are labeled $\pm 1$. The energy widths of the pair of innermost bands, denoted by conduction and valence bands, is also shown.}
\label{fig9}
\end{center}
\end{figure}

Figure~\ref{fig9} gives a general picture of the band flattening in collapsed chiral carbon nanotubes in terms of the energy spans of their bands around the Fermi level.  
The widths of all band sets decrease gradually (with a similar trend) as the moiré angle diminishes independently of the family each tube belongs to. 
This picture neatly illustrates  that a small moiré angle is the only requirement needed to obtain flat bands in all types of collapsed chiral nanotubes. Band spans of the whole central sets are in agreement with those of TBG, taking into account folding effects \cite{nam_lattice_2017}. 

\begin{figure}[h!]
\begin{center}
\includegraphics[width=0.9\columnwidth]{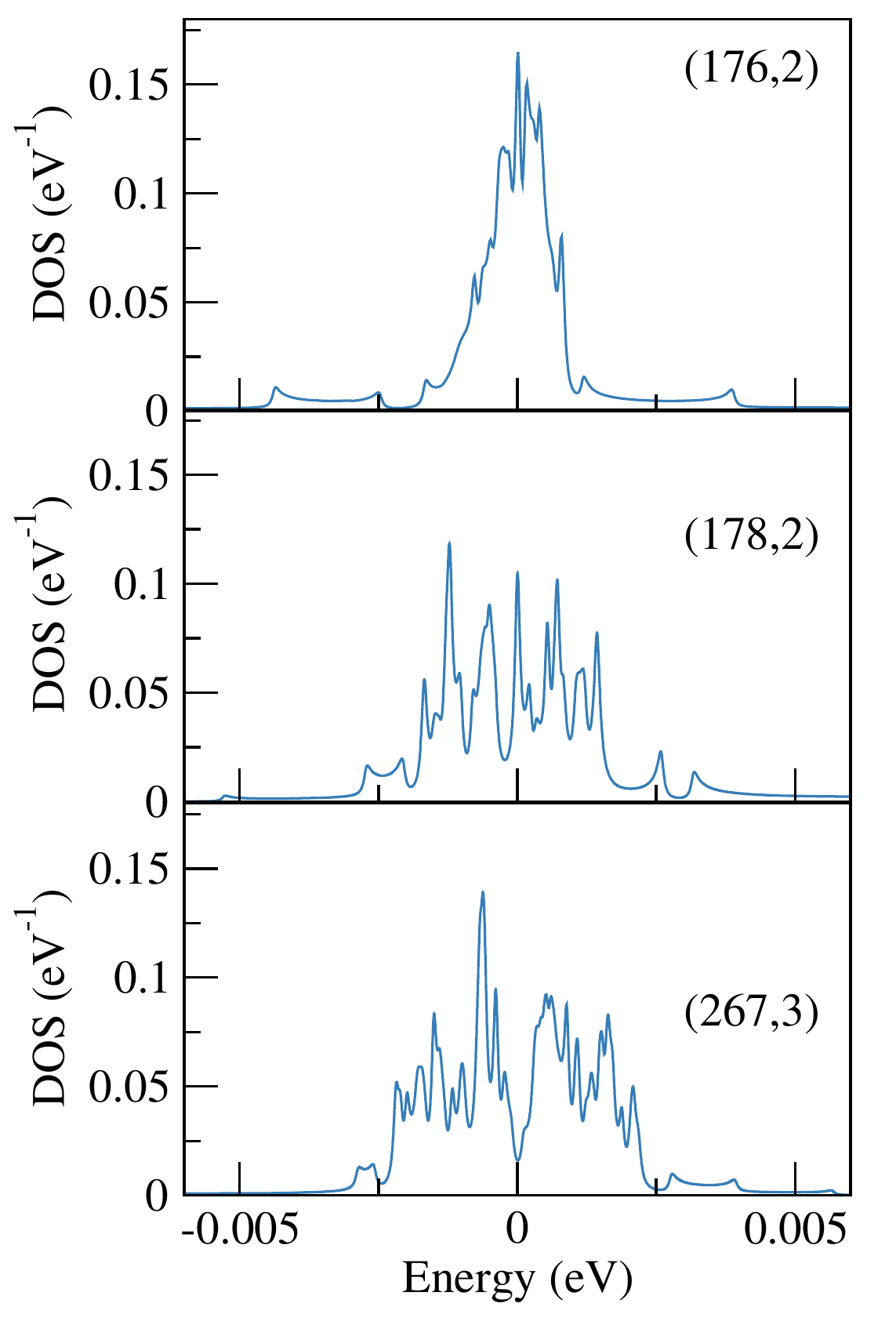}
\caption{From top to bottom: DOS around the Fermi energy of the 2/3-metal nanotube (176, 2) ($\theta_M=1.12\degree$), the semiconductor nanotube (178, 2) ($\theta_M=1.11\degree$) and the  $\Gamma$-metal nanotube (267,3) ($\theta_M=1.11\degree$).}
\label{fig10}
\end{center}
\end{figure}

\begin{table}
\begin{tabular}{ccrcccr}
\hline
 $(n,m)$ & Family & $\theta_{M}$ & $d_{cyl}$(\AA) & $T_{col}$(\AA) & $N_M$ & $N_A$\\ 
 \hline
(36,2)   & SC & 5.36 & 29.004 & 77.565  & 3 & 2744\\
(132,3) & $\Gamma$ & 2.23 & 104.556 & 186.426 & 6 & 23772\\
(90,2)   & SC & 2.18 & 71.270 & 190.603 & 3 &16568 \\
(94,2)   & SC & 2.09 & 74.401 & 198.987 & 3 & 18056\\
(98,2)   & 2/3 & 2.00 & 77.533 & 69.119 & 1 & 6536 \\
(114,2) & SC & 1.73 & 90.060 & 240.864 & 3 & 26456 \\
(122,2) & 2/3 & 1.62 & 96.324 & 85.872 &  1 & 10088 \\
(195,3) & $\Gamma$ & 1.52 & 153.881 & 274.378 & 6 & 51492 \\
(132,2) & SC & 1.49 & 104.152 & 278.562 & 3 & 35384\\
(198,3) & $\Gamma$ & 1.49 & 156.230  & 278.560 & 6 & 53076\\
(162,2) & SC & 1.22 & 127.642 & 341.390 & 3 & 53144\\
(243,3) & $\Gamma$ & 1.22 & 191.464 & 341.414 & 6 & 79716\\
(166,2) & SC & 1.19 & 130.775 & 349.764 & 3 & 55784 \\
(249,3) & $\Gamma$ & 1.19 & 196.163 & 349.772 & 6 & 83676 \\
(258,3) & $\Gamma$ & 1.15 & 203.210 & 362.294 & 6 & 89796 \\
(176,2) & 2/3 & 1.12 & 138.605 & 123.568 & 1 & 20888\\
(178,2) & SC & 1.11 & 140.170 & 374.898 & 3 & 64088\\
(267,3) & $\Gamma$ & 1.11 & 210.256 & 374.870 & 6 & 96132\\
(202,2) & SC & 0.98 & 158.963 & 425.116 & 3 & 82424\\
(208,2) & SC & 0.95 & 163.661 & 437.724 & 3 & 87368 \\
(312,3) & $\Gamma$ & 0.95 & 245.492 & 437.748 & 6 & 131052\\
 \hline
\end{tabular}
\caption{Data of the nanotubes that appear in all figures: family, moiré angle ($\theta_{M}$), diameter of the cylindrical tube ($d_{cyl}$), length of the collapsed unit cell ($T_{col}$), number of moirés ($N_M$) and atoms ($N_A$) per unit cell.}
\label{table1}
\end{table}

With respect to the degree of the localization, Figure~\ref{fig10} displays the best DOS results for the three main types of nanotubes (as mentioned earlier, there is no observed distinction between the two kinds of semiconductors). At the magic angle, the most prominent peak of the 2/3-metal is substantially higher than those of the other cases. The 2/3-metallic tube also has the most centered and narrower distribution of peaks around the Fermi level. Such narrowing is to be expected, since the group of flattest bands is only four here. We already know that the number of bands that undergo this flattening with diminishing chiral angle correlates with the number of AA-regions in the corresponding moiré pattern of a primitive cell. LDOS calculations shows us that the AA-regions are indeed involved in the formation of flat bands. From the 2/3-metallic case \cite{arroyo-gascon_one-dimensional_2020} we learnt that, in particular, one AA-region per primitive cell induces the isolation of a group of eight central bands, with an inner subset of four remarkable flat bands. Our new findings confirm that this flattening  happens irrespectively of the kind of tube, so that the number of isolated bands remains eight and four per AA-region in a primitive cell. Notice that there is no possibility of degeneration in the number of isolated bands, since we are considering primitive cells. Therefore, the AA-regions cannot be completely identical within a unit cell in spite of their striking similarity at first glance. Under this perspective, the increase of the dispersion in the location of narrow peaks in the DOS with increasing number of AA-regions, as shown in Figure~\ref{fig10}, seems fairly natural. Likewise, a bimodal structure of such distribution in the case of the $\Gamma$-metals matches comfortably with the zigzag structure of their moiré patterns. In fact, contrary to the 2/3-metals and semiconductor tubes, in the case of $\Gamma$-metals no band crosses the Fermi level even at the lowest moiré angle explored ($\theta_M=0.95\degree$, (312,3) tube). These facts make the energy span criterion unreliable
to find magic angles since now the flattest bands are not necessarily those closest to the Fermi level.

Notice also that there are pairs of semiconductors and $\Gamma$-metals sharing the same $\theta_M$ (see Fig.\ \ref{fig9}). In particular, that is the case for the (178,2) and (267,3) with $\theta_M=1.11\degree$. Consistently the DOS criterium classifies both cases as ``magic", highlighting again the driving role of the moiré pattern in flat band engineering. 

Multiplication of the DOS by the diameter of the corresponding nanotube does not change substantively the picture shown in Figure~\ref{fig10}: it enhances the 2/3-metallic CNT with respect the two other classes while balancing the performance between the semiconductor and the $\Gamma$-metallic CNTs. At any rate, there are no large differences; eventually, only experiments can tell whether some CNTs are better than others to explore one-dimensional strongly correlated behavior. Our work shows that moiré physics does not require chiral collapsed CNTs of an specific geometry other than being close to the universal magic angle $\sim 1.1\degree$.
 
Finally, for the reader's convenience, we have collected in Table \ref{table1} the most relevant parameters characterizing all the nanotubes addressed in this work.


\section{Conclusions}
We have analyzed the potential of all families of collapsed carbon nanotubes in flat-band engineering. The rotational symmetries of the uncollapsed tubes determine the structure of the moiré patterns. Thus, in their patterns, 2/3-metals show one AA-region per primitive cell, semiconductors have three in a linear arrangement, and $\Gamma$-metals present six in a zigzag disposition. Remarkably, the three families of collapsed CNTs display flat bands, high density of states and localization in the AA regions of the moiré when 
$\theta_M$ 
decreases. The number of highly flattened bands shows a perfect fourfold proportionality with the number of AA-regions per primitive cell, but otherwise, there are not significative differences. We have explored the criteria employed to find magic angles in moir\'e systems, namely, zero velocity, band span and high DOS, which were shown to yield the same result for 2/3 metals and TBG \cite{arroyo-gascon_one-dimensional_2020}. We have concluded that the DOS criterion is best suited to distinguish magic angle behavior in collapsed chiral CNTs of any type. In particular, the effects are maximized at the same magic moiré angle, namely $\sim 1.1 \degree$.

Our results show that moiré physics is universal, i.e., it will appear in all kinds of chiral collapsed nanotubes, provided that their chiral angle is small enough to host AA-regions with localized states. This provides a new reliable perspective to approach other physical issues related to chiral collapsed CNTs, such as the twisting deformations of collapsed carbon nanotubes \cite{barzegar_spontaneous_2017} and its potential relation with magic angles. We expect our conclusions to stir the experimental search for these one-dimensional, highly correlated systems.

\section*{Acknowledgments}
We thank Gloria Platero for generously sharing her computational resources and Sergio Bravo for helpful discussions. We also thank the Centro de Supercomputaci\'on de Galicia, CESGA, (www.cesga.es, Santiago de Compostela, Spain) for providing access to their supercomputing facilities. This work was supported by grant PID2019-106820RB-C21 funded by MCIN/AEI/ 10.13039/501100011033/ and by “ERDF A way of making Europe”, by grant TED2021-129457B-I0 funded by MCIN/AEI/10.13039/501100011033/ and by the “European Union NextGenerationEU/PRTR” and by grant PRE2019-088874 funded by MCIN/AEI/10.13039/501100011033 and by “ESF Investing in your future”. ESM acknowledges financial support from FONDECYT Regular 1221301, and LC gratefully acknowledges the support from Comunidad de Madrid (Spain) under the Multiannual Agreement with Universidad Complutense de Madrid, Program of Excellence of University Professors, in the context of the V Plan Regional de Investigación Científica e Innovación Tecnológica (PRICIT).

\bibliography{ref}

\end{document}